\documentclass[preprint,superscriptaddress]{revtex4}

\usepackage{amssymb}
\usepackage{amsmath}
\usepackage{amsfonts}
\usepackage{latexsym}
\usepackage{graphicx}

\newcommand{\half}{{\textstyle{1\over 2}}}
\newcommand{\be}{\begin{equation}}
\newcommand{\ee}{\end{equation}}
\newcommand{\bea}{\begin{eqnarray}}
\newcommand{\nn}{\nonumber}
\newcommand{\eea}{\end{eqnarray}}
\newcommand{\R}{r_{h}}
\newcommand{\T}{T_H}
\newcommand{\Sn}{\mathbb{S}}
\begin{document}

\title{Gravitational anomalies: a recipe for Hawking radiation\footnote{This essay received an ``Honorable Mention" in the 2007
Essay Competition of the Gravity Research Foundation.}}

\author{Saurya Das}
\email{ saurya.das@uleth.ca}
\affiliation{Department of Physics\\
 University of Lethbridge\\
 4401 University Drive, Lethbridge\\
 Alberta -- T1K 3M4, Canada}

\author{Sean P. Robinson}
\email{spatrick@mit.edu}
\affiliation{Department of Physics\\
 Massachusetts Institute of Technology\\
 Cambridge, Massachusetts 02139, USA}

\author{Elias C. Vagenas}
\email{evagenas@phys.uoa.gr}
\affiliation{Nuclear and Particle Physics Section\\
 Physics Department\\
 University of Athens\\
 GR-15771, Athens, Greece}
\affiliation{Research Center for Astronomy \& Applied Mathematics\\
 Academy of Athens\\
 Soranou Efessiou 4\\
 GR-11527, Athens, Greece}

%\date{}

\preprint{arXiv:0705.2233v3}

\begin{abstract}
We explore the method of Robinson and Wilczek for deriving the
Hawking temperature of a black hole. In this method, the Hawking
radiation restores general covariance in an effective theory of
near-horizon physics which otherwise exhibits a gravitational
anomaly at the quantum level. The method has been shown to work
for broad classes of black holes in arbitrary spacetime
dimensions. These include static black holes, accreting or
evaporating black holes, charged black holes, rotating black
holes, and even black rings. In the cases of charged and rotating black holes, the expected
super-radiant current is also reproduced.
%An interpretation of
%Hawking radiation as an anomalously driven current may be useful
%in the understanding of the black hole information loss problem.
\end{abstract}

\maketitle

%\baselineskip=18pt

%Black holes which are large compared to the Planck scale will most certainly Hawking radiate and shed their mass
%\cite{Hawking:1974sw}, and the radiation may even be detectable in
%colliders such as the Large Hadron Collider \cite{LHC}. If they
%continue to radiate however, the final state being simply thermal
%radiation, the bulk of information that entered the horizon at
%early times is lost forever. This is the so-called {\it
%information loss problem}.

%Information loss results from the evolution of a pure density
%matrix into a mixed one. Since the latter cannot take place in the
%course of unitary evolution, whether one should regard such loss
%as a paradox or not depends on whether or not one believes that
%quantum mechanics will hold in the presence of black holes. In any
%case, a better understanding of Hawking radiation from fundamental
%physical principles may shed more light on the problem.

In this essay, we will review a method developed by Robinson and
Wilczek \cite{Robinson:2005pd}, and subsequently by others
\cite{phd, Iso:2006wa, Iso:2006ut, Murata:2006pt, Vagenas:2006qb,
Setare:2006hq, Xu:2006tq, Iso:2006xj, Jiang:2007gc, Jiang:2007pn,
Jiang:2007wj, Iso:2007kt, Kui:2007dy, Shin:2007gz, Peng:2007pk,
Jiang:2007mi, Chen:2007pp, Miyamoto:2007ue, Iso:2007hd,
Jiang:2007pe}, in which Hawking radiation \cite{Hawking:1974sw}
enters the theory as a mechanism that cancels a would-be
gravitational anomaly, thus saving the unitarity of quantum
mechanics near a black hole from potential violation.
%A careful study of the evolution of such a radiation flux from
%very early to very late times may indeed provide a better
%understanding of the information loss problem \cite{ILP}, but we
%will have little more to say about the information loss problem
%here.
%
%While various approaches of quantum gravity
%(such as String Theory and Loop Quantum Gravity) attempt to do just that,
%they are often accompanied by radical departures from commonly held notions of
%spacetime, not to mention various unproven (and sometimes perhaps
%un-provable) assumptions. In this essay, we show on the contrary,
%that Hawking radiation is closely tied to a
%well understood physical mechanism - that of quantum anomalies,
%which signify the breakdown of a classical symmetry on quantization.
%More precisely, to restore the symmetry of general covariance
%for quantum fields in a black hole background, a Hawking radiation
%flux of the field is both necessary and sufficient \cite{Robinson:2005pd,phd}!
%
For purposes of illustration, we will first show the method for
the simplest case, following \cite{Robinson:2005pd}, and then
describe subsequent developments. Consider a $d$-dimensional,
static, spherically symmetric black hole spacetime
 with metric:
\be ds^2 = -f(r)dt^2 +\frac{dr^2}{f(r)} +r^2
d\Omega^2_{(d-2)},\label{metric1} \ee
where $d\Omega_{(d-2)}$ is the line element on $\Sn^{(d-2)}$ and
$f(r)$
 is an arbitrary
%, but suitably well behaved,
function of the radial coordinate  $r$ only.
% This radial coordinate is the ``geometric'' radius, in the sense that the
%area of surface of constant $t$ and $r$ is $\Vol (\Sn^{(d-2)} )r^{(d-2)}$.
The more general case where $g_{tt} \neq g^{rr}$ is considered in
\cite{phd,Vagenas:2006qb}.
% When studying black hole spacetimes, we generally assume asymptotic flatness
% --- that is $\mathrm{lim}_{r\to\infty}f(r)=1$ --- but this is not essential
%to the analysis.
For a black hole, we require the existence of a horizon at
coordinate $r=\R$ determined by the condition $f(\R)=0$. The
surface gravity of the horizon is given by
$\kappa\equiv\half\partial_rf(r)\Big|_{r=\R}$. For this simple
spacetime, the varying definitions of ``horizon'' (event, Killing,
{\it etc.}) all coincide. For a discussion of the role of horizon
taxonomy in black hole thermodynamics, see \cite{Visser:2001kq}.
%For example, $r=\R$ is an event horizon (separates trajectories which can
%reach infinity from those that fall into the central singularity), an
%apparent horizon (bounds a region where the areas of a sphere of outgoing
%null rays decreases with time, rather than increases), and a Killing horizon
%(a surface where the norm a Killing vector vanishes). The Killing vector on
%interest here is the generator of translations in $t$. In more general
%spacetimes, the Killing horizon condition is the important one, but as
%discussed in \cite{phd}, the method used here requires that  for static
%spherically symmetric metrics, Killing and event horizons coincide.
 Because the horizon is a Killing horizon for translations in $t$, the
associated conserved quantity fails to behave like a proper energy
inside the horizon where $\partial_t$ becomes spacelike. As a
symptom of this pathology, the vacuum state constructed to have
zero Fock space occupation number with respect to $\partial_t$
(the so-called Boulware state \cite{Boulware:1974dm}) contains
divergences due to a pile up of outgoing
 high frequency modes at the horizon.

The essence of the Robinson-Wilczek method is to take the lessons
of effective field theory seriously: the physics observed by a
given experiment should be describable by an effective theory of
only those degrees of freedom accessible to the experiment. The
effective theory is derived from the fundamental theory by
integrating out inaccessible degrees of freedom.
%In Minkowski space quantum field theory, the limiting factor on accessibility
% is typically the available energy of the experimental probe. In a black hole
%spacetime, however, the situation is more complicated.
In a black hole spacetime, degrees of freedom inside the horizon
are inaccessible to the outside observer. Outgoing near-horizon
modes are also inaccessible due to their diverging energy, as
discussed above. To form the effective theory of the outside
observer, we must remove these modes from all fields in the
theory.
%(See Figure \ref{fig1}.)
%\begin{figure}
%\includegraphics{penrose.eps}
%\caption{\label{fig1} A piece of the Penrose diagram of a black
%hole spacetime
 %is shown with a microscopic view of a patch near the horizon. The dashed
%arrows represent unoccupied modes; solid arrows represent occupied
%modes. The
 %white area is the infinitesimal region near the horizon where the effective
 %physics lacks outgoing modes. Figure borrowed from \cite{Robinson:2005pd}.}
%\end{figure}
Upon doing so, however, we will encounter a problem. Arguments that theories of quantum gravity should be formulated in terms of the effective degrees of freedom accessible to a given observer have also appeared in \cite[for example]{Padmanabhan:2003ub, Padmanabhan:2003gd}.

As a probe of this background geometry, we consider
%for the remainder of this essay
a scalar field with arbitrary self-interactions. By expanding the
field in partial wave modes using $(d-2)$-dimensional spherical
harmonics and taking the near-horizon limit, we see that the
action for each partial wave mode reduces to that of a free,
massless, $(1+1)$-dimensional scalar field on the $r$-$t$ section
of the original spacetime. The only remnant of $d$-dimensional
physics is the degenerate angular momentum
 quantum numbers, which are now just labels on otherwise identical fields. This effective
dimensional reduction is demonstrated explicitly in
 \cite{phd}, but has been seen in earlier work \cite{Govindarajan:2000ag, Camblong:2004ye}.

 Since we have eliminated the outgoing modes, the effective near-horizon
theory is chiral. As shown in \cite{louis}, $(1+1)$-dimensional
chiral theories exhibit a gravitational anomaly and therefore fail
to covariantly conserve the energy-momentum tensor. For the case
of an ingoing scalar field, the anomaly takes the form
\cite{bert2}:
\be \nabla_{\mu}T_{\nu}^{\mu} = \frac{1}{\sqrt{-g}}\partial
_{\mu}N^{\mu}_{\nu}\equiv A_{\nu}, \label{anomaly} \ee
where \be
N^{\mu}_{\nu}=\frac{1}{96\pi}\varepsilon^{\beta\mu}\partial_\alpha
\Gamma^{\alpha}_{\nu\beta}, \label{Nquantity} \ee
$\varepsilon^{\mu\nu}$ is the anti-symmetric unit tensor
($\varepsilon^{01}=1$), and $\Gamma^{\alpha}_{\nu\beta}$ is the
Christoffel connection on the $(1+1)$-dimensional spacetime.
Equation (\ref{anomaly}) can be solved as
\bea
T^{t}_{t}&=&-\frac{(K+Q)}{f}-\frac{B(r)}{f}-\frac{I(r)}{f}+T^{\alpha}_{\alpha}(r), \\
T^{r}_{r}&=&\frac{(K+Q)}{f}+\frac{B(r)}{f}+\frac{I(r)}{f},\label{emcpts1}\\
T^{r}_{t}&=&-K+C(r)=-f^{2}T^{t}_{r}, \eea
where \bea
B(r)&\equiv& \int^{r}_{\R}f(x)A_{r}(x)dx, \\
C(r)&\equiv& \int^{r}_{\R}A_{t}(x)dx, \\
I(r)&\equiv&\frac{1}{2}\int^{r}_{\R}T^{\alpha}_{\alpha}(x)f'(x)dx.
\eea
The constants $K$, $Q$, and the trace $T^\alpha_\alpha(r)$ are
undetermined.

However, Equation (\ref{anomaly}) does not hold over the entire
spacetime. In constructing the effective theory, we imposed
this condition only at the horizon. Equation (\ref{anomaly}) will
hold only in an infinitesimal region about the horizon between
$\R\pm\epsilon$ in the limit $\epsilon\rightarrow 0$. Since the
fundamental theory is generally covariant, the quantum effective
action $W$ must be invariant under a coordinate transformation
with parameter $\lambda^\nu$: \be \lim_{\epsilon\rightarrow
0}\delta_{\lambda}W=0. \ee Performing this coordinate variation
explicitly, we find that
\bea -\delta_\lambda W &=&\int d^{2}x
\sqrt{-g}\lambda^{\nu}\nabla_{\mu}\left\{T^{\mu}_{i\,\nu}\Theta_{+}
+ T^{\mu}_{o\,\nu}\Theta_{-} + T^{\mu}_{\chi\,\nu}H\right\}
\nn \\
&=&\int d^{2}x
\sqrt{-g}\lambda^{t}\left\{\partial_{r}\left(N^{r}_{t}H\right)+
\left(T^{r}_{o\,t}-T^{r}_{\chi\,t}+N^{r}_{t}\right)\partial_r\Theta_{+}
+
\left(T^{r}_{i\,t}-T^{r}_{\chi\,t}+N^{r}_{t}\right)\partial_r\Theta_{-}\right\}\nn\\
&&+\int d^{2}x \sqrt{-g}\lambda^{r}
\left\{\left(T^{r}_{o\,r}-T^{r}_{\chi\,r}\right)\partial_r\Theta_{+}
+
\left(T^{r}_{i\,r}-T^{r}_{\chi\,r}\right)\partial_r\Theta_{-}\right\}.
\label{vareffaction2} \eea
We have written the total energy-momentum tensor as the sum of
{\emph inside}, {\emph outside} and {\emph chiral} parts:
\be T^{\mu}_{\nu}=T^{\mu}_{i\,\nu}\Theta_{+} +
T^{\mu}_{o\,\nu}\Theta_{-} + T^{\mu}_{\chi\,\nu}H,
\label{emtensor} \ee where $\Theta_{\pm}\equiv\Theta\left(\pm
(r-\R)-\epsilon\right)$ are step functions and $H=1-
\Theta_{+}-\Theta_{-}$ is a
 ``top hat'' function which is equal to unity between
$\R\pm \epsilon$ and zero elsewhere. The quantities
$T^{\mu}_{i\,\nu}$ and $T^{\mu}_{o\,\nu}$ are covariantly
conserved inside and outside the horizon, respectively. However,
$T^{\mu}_{\chi\,\nu}$  is not covariantly conserved and expresses
the  anomalous chiral physics on the horizon.

Taking derivatives of the $\Theta$ functions and expanding for
small $\epsilon$, we find that
\bea \delta_{\lambda}W&=& \int
d^{2}x\lambda^{t}\left\{\left[K_{o}-K_{i}\right]\delta\left( r-\R
\right)
- \epsilon\left[K_{o}+K_{i}-2K_{\chi}-2N^{r}_{t}\right]\partial_r\delta\left( r-\R \right)+\ldots\right\}\nn\\
&&- \int d^{2}x\lambda^{r}\left\{\left[\frac{K_{o} +Q_{o}+K_{i}+Q_{i}-2K_{\chi}-2Q_{\chi}}{f}\right]\right.\nn\\
&&-\left.\epsilon\left[\frac{K_{o}
+Q_{o}-K_{i}-Q_{i}}{f}\right]\partial_r\delta\left(r-\R\right)+\ldots\right\}
\label{vareffaction3}. \eea It is easily seen in Equation
(\ref{vareffaction3}) that only the on-horizon values of the
energy-momentum tensor will contribute to the gravitational
anomaly. Since the parameters $\lambda^{t}$ and $\lambda^{r}$ are
independent, each of the four terms in square brackets in Equation
(\ref{vareffaction3}) must vanish simultaneously, but only needs to do
so on the horizon.
%Being careful about signs as we approach the
%horizon from above and below,
These conditions yield \bea
K_{o}=K_{i}=K_{\chi}+\Phi,\\
Q_{o}=Q_{i}=Q_{\chi}-\Phi, \label{conditions} \eea where \be
\Phi=N^{r}_{t}\Big|_{\R}=\frac{\kappa^2}{48\pi}= \frac{\pi}{12}
\T^2 \label{phi} \ee
and $\T$ is the Hawking temperature \be \T=\frac{\kappa}{2\pi}.
\label{hawktemp} \ee
%The result of demanding anomaly cancelation
%is a boundary condition on the flux components of the
%energy-momentum tensor at the horizon
% --- with the bulk properties determined by the conservation equation ---
%which fixes the magnitude of the flux $\Phi$ as in Equation (\ref{phi}).
Equation (\ref{phi}) is exactly the flux per partial mode that
would result from a thermal distribution at the Hawking
temperature in the full $d$-dimensional theory. That is, this flux
is necessary and sufficient to restore general covariance at the
quantum level, although we have not shown that the full spectrum
is in fact thermal.

%
%That the above can be
%easily extended to time-dependent and various
%other black holes demonstrates the robustness of the
%mechanism \cite{Vagenas:2006qb}. We elaborate on the above in the rest
%of the essay.

The above construction was extended to the most general static
spherically symmetric metrics in \cite{phd, Vagenas:2006qb}. The
method was also successfully implemented in a number of special
cases \cite{Setare:2006hq, Jiang:2007pn, Jiang:2007wj, Kui:2007dy,
Shin:2007gz, Peng:2007pk, Jiang:2007mi, Jiang:2007pe}. Notable
among the special cases are \cite{Peng:2007pk}, where the
spacetime in question exhibits a global deficit solid angle, and
\cite{Jiang:2007mi, Jiang:2007pe}, which study radiation from
cosmological horizons (as opposed to black hole horizons) in de
Sitter black hole spacetimes.  Moreover, the time-dependent
spherically symmetric spacetime, which includes evaporating and
accreting black holes, was also studied in \cite{Vagenas:2006qb}
by use of the Vaidya metric. In this dynamical spacetime,
deviations from the purely thermal (blackbody) flux were derived
as expected. To our knowledge, this is the only direct calculation
of the Hawking flux per partial wave in a time dependent case.

An important generalization of the method was presented by Iso,
Umetsu, and Wilczek \cite{Iso:2006wa} to charged black holes by
using the gauge anomaly as well as the gravitational anomaly. This
was further extended to four-dimensional Kerr-Newman black holes
in \cite{Iso:2006ut, Murata:2006pt}. In these cases of charged and
rotating black holes, we must take into account the energy flow
and the super-radiant gauge currents.
%Thus, the gravitational as well as the gauge anomalies are considered.
Despite the lack of spherical symmetry in the case of
four-dimensional Kerr black holes, the essential ingredient of
reduction to an $r$-$t$ theory still
works. %In this theory,
The angular isometry generates an effective $U(1)$ gauge field in
the $1+1$ theory, with the $m$ quantum number serving as the charge
of each partial wave. At this point the analysis of
\cite{Iso:2006wa} goes through, and the known result is obtained
with angular momentum acting like a chemical potential for the
effective charge.

Furthermore, the method was extended to $d$ dimensions in
\cite{Xu:2006tq, Iso:2006xj, Jiang:2007gc} for Myers-Perry and
Myers-Perry-(A)dS black holes. In this case, each independent
angular momentum becomes a factor in a $U(1)^N$ product gauge
group. Then the previous analysis goes through.

We have described the Robinson-Wilczek method above as described
in \cite{Robinson:2005pd}:
 outgoing modes are eliminated only near the horizon. We should
stress that some authors \cite[for example]{Iso:2006wa} eliminate
modes only outside the horizon. Moreover, they eliminate the
ingoing modes, which are irrelevant at the classical level to
physics outside the horizon. While this makes no difference in the
simple case described here, using only the ingoing exterior modes
was found to be essential when either gauge symmetries or rotation
are present,
as explained in \cite{Iso:2006xj}.\\

%conclusions
In this essay, we have presented a view of Hawking radiation in
which it is a consequence of cancelation of gravitational
anomalies that arise from following the philosophy of effective
field theory.
%Hence, if we
%adopt the point of view that nature dislikes anomalies, and that they
%should not be present in any final theory (in this case, Quantum Gravity),
%then Hawking radiation is inevitable!
%Then being such a
%fundamental effect, it is quite plausible that the problem of
%information loss that seems to accompany Hawking radiation is an
%artifact of the various assumptions that are made in the traditional
%derivation of Hawking radiation, and that no such problem exists in
%the current picture.
%Clearly, more work needs to be done in this direction,
%on which we hope to report in the near future.
The broad successes of this approach, as outlined above, are sufficient to declare it legitimate, but open questions remain:\\
1) The method has not been applied to black objects of non-spherical topology.\\
2) No proof exists that anomaly cancelation induces a thermal radiation spectrum, although important steps have been taken in this direction \cite{Iso:2007kt, Iso:2007hd}.\\
3) We constructed the effective theory ``by hand''. Could it be constructed directly by integrating out modes in the path integral?\\
4) The method has remarkable qualitative similarity to the near-horizon conformal field theory approaches used in \cite{Govindarajan:2000ag, Camblong:2004ye, Strominger:1997eq, Carlip:1998wz, Solodukhin:1998tc}. What is the quantitative connection?\\
%There is an obvious synergy, or at least a strong analogy, between
% the program of \cite{Govindarajan:2000ag, Camblong:2004ye, Strominger:1997eq, Carlip:1998wz, Solodukhin:1998tc} and the one discussed here, but making this
%connection explicit remains an open problem. In those contexts, the
%dimensional reduction is used in the context of showing
% that the near-horizon physics is not only effectively two dimensional, but in
% fact governed by $SL(2,\mathbb{R})$ conformal field theory, which
%can then be used to compute the entropy. In these works, there is a partial
%breaking of diffeomorphism invariance at the horizon which is critical to the
% formalism. In both programs, the effective reduction of physics to $1+1$
%physics is an essential ingredient, but to our knowledge it has only been
%shown explicitly for specific classes and special cases of black hole metrics.
% The examples that have been studied are very broad, but not exhaustive. Some
%of the remaining open cases, in particular black rings, could prove very
%difficult to show. A general proof that spacetime kinematics near any event
%horizon (or perhaps Killing horizon) forces physics to be effectively
%two-dimensional there would be a major advance.\\
%
5) The method uses only spacetime kinematics, not dynamics. Thus
it seems unlikely that it can be used to calculate black hole
entropy, but the similarities to the methods of
\cite{Govindarajan:2000ag, Camblong:2004ye, Strominger:1997eq,
Carlip:1998wz, Solodukhin:1998tc}, which can calculate entropy via
the Cardy formula, leave this
possibility open.\\

\emph{Note added}: References \cite{Chen:2007pp, Miyamoto:2007ue} appeared shortly after completing this manuscript, addressing the first open question listed above. The Robinson-Wilczek method has now been successfully applied to five-dimensional black ring spacetimes, showing that the method continues to work for black objects of non-spherical horizon topology. Again, the near-horizon physics reduces to a $1+1$ theory with a $U(1)$ gauge symmetry arising from the ring's angular momentum. This work is significant because the nonseparable coordinates typical to black ring spacetimes have impeded previous detailed study of their thermodynamics. The simplicity inherent to near-horizon physics as employed in the Robinson-Wilczek method may lead to further advances in black ring thermodynamics.\\

%%%%%%%%%%%%%%%%%%%%%%%%%%%%%%%%%%%%%%%%%%%%%%%%%%%%%%%%%%%%%%%%%%%%%%%%%%%%%%%%%%%%%%%%%%%%%%%%%%%%%%
%%%%%%%%%%%%%%%%%%%%%%%%%%%%%%%%%%%%%%%%%%%%%%%%%%%%%%%%%%%%%%%%%%%%%%%%%%%%%%%%%%%%%%%%%%%%%%%%%%%%%%
%%%%%%%%%%%%%%%%%%%%%%%%%%%%%%%%%%%%%%%%%%%%%%%%%%%%%%%%%%%%%%%%%%%%%%%%%%%%%%%%%%%%%%%%%%%%%%%%%%%%%%
%%%%%%%%%%%%%%%%%%%%%%%%%%%%%%%%%%%%%%%%%%%%%%%%%%%%%%%%%%%%%%%%%%%%%%%%%%%%%%%%%%%%%%%%%%%%%%%%%%%%%%
\begin{acknowledgements}
 E.~C.~V.~is supported by the Greek State Scholarship Foundation
(IKY). The work of S.~D.~was supported by the Natural Sciences and
Engineering Research Council of Canada and by the Perimeter
Institute for Theoretical Physics.
\end{acknowledgements}
%%%%%%%%%%%%%%%%%%%%%%%%%%%%%%%%%%%%%%%%%%%%%%%%%%%%%%%%%%%%%%%%%%%%%%%%%%%%%%%%%%%%%%%%%%%%%%%%%%%%%%
%%%%%%%%%%%%%%%%%%%%%%%%%%%%%%%%%%%%%%%%%%%%%%%%%%%%%%%%%%%%%%%%%%%%%%%%%%%%%%%%%%%%%%%%%%%%%%%%%%%%%%
%%%%%%%%%%%%%%%%%%%%%%%%%%%%%%% BIBLIOGRAPHY %%%%%%%%%%%%%%%%%%%%%%%%%%%%%%%%%%%%%%%%%%%%%%%%%%%%%%%%%

%%%%%%%%%%%%%%%%%%%%%%%%%%%%%%%%%%%%%%%%%%%%%%5
\end{document}